# Inverse magnetic hysteresis of the Josephson supercurrent: study of the magnetic properties of thin niobium/permalloy (Fe$_{20}$Ni$_{80}$) interfaces

Roberta Satariano,[1] Loredana Parlato,[1,2] Antonio Vettoliere,[3] Roberta Caruso,[1] Halima Giovanna Ahmad,[1,2] Alessandro Miano,[1] Luigi Di Palma,[1] Daniela Salvoni,[2,4] Domenico Montemurro,[1,2] Carmine Granata,[3] Gianrico Lamura,[5] Francesco Tafuri,[1] Giovanni Piero Pepe,[1,2] Davide Massarotti,[6] and Giovanni Ausanio[1,2]

[1]*Dipartimento di Fisica "Ettore Pancini," Università degli Studi di Napoli "Federico II," c/o Complesso di Monte Sant' Angelo, via Cintia 21, I-80126 Napoli, Italy*
[2]*Consiglio Nazionale delle Ricerche – SPIN, c/o Complesso di Monte Sant' Angelo, via Cintia 21, I-80126 Napoli, Italy*
[3]*Consiglio Nazionale delle Ricerche – ISASI, c/o Comprensorio Olivetti, via Campi Flegrei 34, I-80078 Pozzuoli, Italy*
[4] *Dipartimento di Ingegneria Chimica, dei Materiali e delle Produzioni Industriali, Università degli Studi di Napoli "Federico II," Piazzale Tecchio 80, 80125 Napoli, Italy*
[5] *Consiglio Nazionale delle Ricerche – SPIN, Corso Ferdinando Maria Perrone 24, I-16152 Genova, Italy*
[6]*Dipartimento di Ingegneria Elettrica e delle Tecnologie dell'Informazione, Università degli Studi di Napoli "Federico II," via Claudio 21, I-80125 Napoli, Italy*

We propose a picture for the magnetic properties of superconductor/ferromagnet (S/F) heterostructures based on Nb and permalloy (Py: Fe$_{20}$Ni$_{80}$). By measuring the magnetic moment as a function of the temperature in S/F/S trilayers for different thicknesses of the middle F layer, we give evidence of the presence of a magnetic stray field of the F layer. For values of F-layer thickness below a threshold, we establish a correlation between the magnetic measurements of the S/F/S trilayers and the anomalous magnetic dependence of the critical current in S/insulator/thin superconducting film/F/S (SIsFS) Josephson junctions (JJs). These complementary investigations provide a self-consistent method to fully characterize S/F heterostructures and possibly demonstrate effects arising from the mutual interactions between ferromagnetism and superconductivity. A shift in the Fraunhofer critical current oscillations has been observed in the opposite direction to the one commonly observed in JJs with F barriers, as it has been recently predicted by inverse and electromagnetic proximity theories. This inverse memory effect is relevant for the design of these heterostructures as memory cells and spintronic devices.

## I. INTRODUCTION

Superconductivity and ferromagnetism can coexist in Fe-based superconductors [1-3] and superconductor/ferromagnet (S/F) heterostructures [4,5]. However, the former are still research topics and far from being used in applications. On the other side, S/F heterostructures have gained much interest because of their unique properties from both theoretical and technological points of view. Superconducting correlations can be induced, as in normal metal [6], also into the F layer by Cooper pairs migrating from the adjacent S over a typical distance, i.e., the coherence length in the F layer $\xi_F$, of the order of 1-10 nm. This phenomenon is usually referred to as proximity effect. Due to the exchange splitting in F, the Cooper pair gains a nonzero momentum, resulting in an oscillating pairing wave function in the F layer, leading to peculiar phenomena, such as the oscillatory superconducting transition temperature $T_c$ [7], the spatial oscillation of electronic density of states [8,9], the suppression of Andreev reflection [10], and the $\pi$ Josephson junctions (JJs) [11]. More recently, the so-called spin triplet superconductivity in S/F heterostructures with inhomogeneous magnetization $M$ has opened a research field called superconducting spintronics [12-14]. Moreover, there are a lot of potential applications in artificial structures consisting of S/F multilayers. S/F hybrids are optimal systems to be used as ultrafast superconducting optical detectors and eventually as single photon detectors [15,16]. Furthermore, the possibility to use SFS JJs as random access memory has been demonstrate [17], while the demonstration of tunnel JJs S/insulator/thin superconducting film/F/S (SIsFS) -based memory elements with characteristic voltage levels compatible with standard single flux quantum circuitry has also been reported [18-20]. A deeper understanding of the magnetization of the F layer ($M_F$) and of the S/F multilayers, also in the presence of an external magnetic field $H$, is thus a key requirement for further advances in the field, to establish a closer correlation between the magnetic and superconducting properties of these devices and the actual mechanisms which rule the mutual influence between the S and F layers.

There can be different ways to modify the magnetic properties in S/F structures by the leakage of the induction field from the F to the S layer. First, the stray field of the F layer can penetrate the S and induce screening currents [21]. The inverse proximity effect, namely, the transfer of the magnetic moment $m$ from the F to the S subsystem is also possible [22,23]. This phenomenon is related to the Cooper singlet pairs localized in proximity to the S/F interface: an electron with the spin aligned along the exchange field can easily penetrate the F layer, while an electron with the opposite spin tends to stay in the S layer. As result, the surface of the S layer down to a depth of the order of the Cooper pair size, i.e., the superconducting coherence length $\xi_S$ ($\sim 1 - 100\ nm$) acquires a net magnetization $M_{SC}$ with opposite direction to $M_F$. Recently, it has been proposed that, neglecting this short-range inverse proximity effect and the stray fields of the F layer, the direct S/F proximity effect is always responsible for the generation of screening supercurrents in response to the presence of a vector potential at the S/F interface, the so-called electromagnetic proximity effect [24]. These Meissner supercurrents generate a magnetic induction field $B_{SC}$ in the S film which is antiparallel to $M_F$ and decays at distances of the order of the London penetration depth $\lambda_L$.

In the context of S/F hybrid structures based on niobium (Nb) and permalloy (Py: $Ni_{80}Fe_{20}$), often used in spintronic devices, there is rich literature concerning the effect of the stray fields of Py films with striped domains or out-of-plane magnetization on the Nb films, e.g., on the vortex lattice [25-28]. Nevertheless, the role of the stray field generated by ultrathin ferromagnetic film with an in-plane magnetization deserves more experimental investigations. In this paper, we explore the in-plane magnetization effects arising in hybrid structures with Py as F layers with thickness $d_F$ of the order of 1-10 nm.

By performing systematic magnetic measurements on Nb/Py/Nb trilayers with different $d_F$, we show that, only above a $d_F$ threshold value (>7 nm), the stray field can penetrate the adjacent S layers, producing relevant effects. Indeed, we have direct evidence of F stray fields through the measurement of the temperature dependence of the remanent magnetic moment $m_r$ and of the hysteresis loop across $T_c$. For values of $d_F$ thickness lower than such a threshold, the effects of the stray fields become negligible. The comparative study of the zero-field cooling (ZFC) and field cooling (FC) measurements as a function of $d_F$ offers solid self-consistent criteria to explain the anomalies observed in the Josephson magnetic response of JJs with the same bottom layer thickness [20]. We provide evidence of an unconventional magnetic field behavior of the critical current $I_c$ characterized by an inverted magnetic hysteresis, i.e., an inverted shift of the whole magnetic field pattern when sweeping the external field in Josephson tunnel SIsFS junctions with Py thickness of 3 nm. This occurrence has been predicted by both inverse [29] and electromagnetic proximity effect [30]. Here, we report an experimental study: having ruled out the possibility that this uncommon behavior could be related to the stray fields of the F layer by means of our magnetic characterizations, we experimentally show that the Josephson magnetometry [17] can be successfully used to detect the induction field into the S layer due to proximity effects, which represents a powerful tool to investigate spin polarization phenomena in specific junction configurations.

## II.    EXPERIMENT

The Nb (30 nm) /Py ($d_F$) / Nb (400 nm) trilayers were deposited on oxidized Si wafers with lateral dimensions of 10 mm x 10 mm using DC magnetron sputtering in an ultrahigh vacuum chamber. The thickness of the Nb layers was chosen to allow comparison with the properties of our previous realized SIsFS JJs [20]. First, a Nb film having a thickness of 40 nm was deposited on the oxidized Si substrate at a rate of 1.2 $\pm$ 0.1 nm/s, which leads to a margin of error of 10% on the Nb thicknesses. Then, with vacuum breaking, the surface of Nb was cleaned by a soft etching procedure by using an ion gun. During the ion beam etching, the Ar pressure was fixed at 3 mTorr, and a power discharge of 8 W was applied for 9 min to remove ~10 nm of Nb oxide layer, so that the effective thickness of the Nb layer was 30 nm. After the cleaning process, the Nb was moved into an adjacent vacuum system, and a film of Py ranging from 3 to 10 nm was sputtered by a magnetron source at a rate of 0.7 nm/s at room temperature. Since no external field was applied during the deposition, we expect that, for these thicknesses, these films will not exhibit any uniaxial anisotropy [31], but instead an in-plane anisotropy with Néel domain wall [32]. Finally, a 400-nm top Nb layer was deposited by further DC sputtering. With our growth conditions, it is reasonable to expect that the 30-nm Nb layer will have a $T_c$ of ~ 6 K [27], and the 400-nm Nb layer will have a $T_c$ that approaches the bulk value

[33]. The transition of the trilayers in the normal state is observed at $T_c$ = 8.8 K, both in the thermoremanence measurements and in the ZFC and FC curves, as it will be detailed in the following paragraphs. The trilayers were synthesized with different thicknesses of the Py layer: 3, 7, and 10 nm.

The DC magnetic moment *m* of the trilayers was measured using a vibrating sample magnetometer equipped with a He-flow cryostat of Oxford Instruments-MagLab and a commercial DC superconducting squid interference device provided by Quantum Design. All measurements were performed by applying *H* parallel to the film plane. The magnetometers detect the superposed signal from the F and S films. To distinguish the behavior of every component of the heterostructures, the magnetic measurements have been performed below and above $T_c$. Since the paramagnetic signal from the Si substrate and from the Nb films in the normal state are negligible above $T_c$ in the range of the applied field, the magnetic response of the heterostructures is that of the Py layers. The Curie temperature of bulk face-centered cubic Py far exceeds room temperature ($T_{Curie} \cong$ 872 K) [34]. Since the Curie temperature can be frustrated in ferromagnetic ultrathin films [35], we verified experimentally that the Curie temperature of our Py films was > 100 K and that $\mu_0 M_s$ remains almost constant below such a temperature. Hence, we can state that the magnetization of the F-layer is constant in the temperature range of this paper (< 15 K). For this reason, the changes below $T_c$ will be mainly ascribed to the presence of the S layers. Hence, hysteresis loops were measured above $T_c$ (12 K) by sweeping the applied field $\mu_0 H$ from -30 to 30 mT. After cooling the demagnetized sample down to 4 K in zero field [36], hysteresis loops were obtained. Furthermore, magnetic thermoremanence measurements were performed. An in-plane magnetic field *H* strong enough to fully saturate the Py layers was applied, then *H* was removed, and the remanent magnetic moment $m_r$ of the trilayers was measured from 12 to 4 K (down curves) and then back to 12 K (up curves). The ZFC curves were measured while warming up the samples to 12 K after cooling the demagnetized samples to 4 K under zero field, and then, at 4 K, the applied field was set to $\mu_0 H_0$ = 30 mT. The samples were then cooled again to 4 K in the presence of the same initial field, and *m* was recorded by increasing temperature up to 12 K (FC curves). In the *m(T)* measurements, the temperature was changed at a fixed rate of 0.2 K/min.

### III. RESULTS AND DISCUSSION

In Figs 1(a), 1(c), and 1(e), the *m(H)* curves, i.e., the hysteresis loops, of the trilayers with $d_F$ of 3, 7, and 10 nm, respectively, are reported at *T* = 12 and 4 K. At *T* = 12 K, we measured the hysteresis loop of the bare Py layer. We observe a squareness, i.e., the ratio $\frac{m_r}{m_s}$, where $m_s$ is the saturation magnetic moment ($m_s$) that approaches the unity, which is evidence of the in-plane anisotropy of the films [37]. From Fig. 1, we can estimate a value of the saturation magnetization $\mu_0 M_s \approx$ 1T comparable with previously published data [38]. The slight deviation from the bulk value (order of percent) suggests that the dead layer at the S/F interface is almost negligible, which is consistent with the small value of roughness, as shown in the Appendix. At $T = 4$ K, the Nb layers contribute to the overall magnetic moment $m$ of the sample. Assuming that the total magnetic moment of the trilayers is a linear superposition of the F and S magnetic contributions, we isolated the Nb hysteresis loop, shown in Figs. 1(b), 1(d), and 1(f), by taking the difference between the $m(H)$ curves above and below $T_c$ [39].

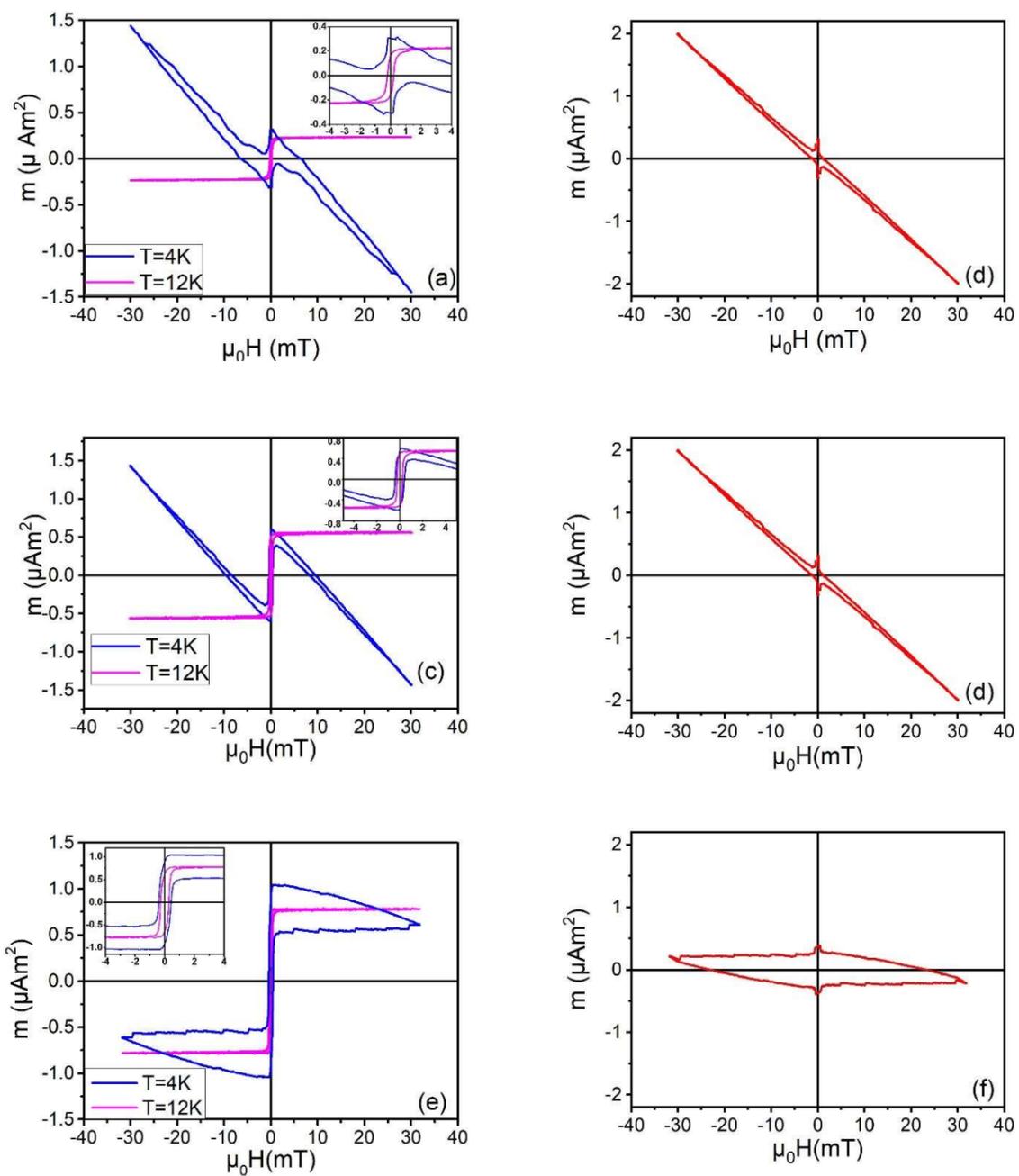

**FIG. 1.** After a standard demagnetizing cycle, the samples were cooled down according the zero-field cooled (ZFC) protocol, and then the hysteresis cycle measurements were performed by applying the maximum field $\mu_0 H = +30 \, mT$. By sweeping the applied field $\mu_0 H$ from +30 to - 30 mT and then back, the hysteresis loops *m(H)* at *T* = 4K (blue) and *T* = 12K (magenta) curves of Nb (30nm) /Py ($d_F$) /Nb (400 nm) trilayers were obtained for different F-thicknesses $d_F$: (a) $d_F$= 3nm, (c) $d_F$= 7 nm, and (e) $d_F$= 10 nm. Insets show a magnification of the m(H) curves between -4 mT and 4 mT. Nb hysteresis loops, isolated by taking the difference between the *m(H)* of Nb (30nm) /Py ($d_F$) / Nb (400 nm) trilayers above and below $T_c$: (b) $d_F$ = 3nm; (d) $d_F$ = 7 nm; (f) $d_F$ = 10 nm.

In Figs. 1(b), 1(d) and 1(f) we remark on the presence of a tiny peak of the magnetization at zero applied magnetic field in all the tested samples. Both the presence of geometrical barriers effects [40] and the reduced superconducting properties ($T_c \cong 6$ K) of the 30 nm-thick Nb layer place it well in the mixed state at *T* = 4 K. This results in the weak opening of the hysteresis cycles as well. Furthermore, small variations of the diamagnetic signal in Figs. 1(b) and 1(d) can be ascribed to the small margin of error on the Nb thickness. In this geometry, the effects of the stray fields should be minimized due to the negligible demagnetizing factor. Nevertheless, the global superconducting behavior in the trilayers with the thickest Py layer in Figs. 1(e) and 1(f) seems determined not only by the applied field but also by the small stray fields penetrating the S layers. Indeed, the presence of the F-stray field strongly reduces the diamagnetic response of the 400 nm Nb layer and drives it in the mixed state as it is proved by the opening of the hysteresis cycle, whose closed area is more than three times greater than the ones evidenced in Figs. 1(b) and 1(d). To further confirm this evidence, we performed a measurement of both magnetic thermoremanence and ZFC-FC temperature dependence of the total magnetic moment.

Figure 2 shows the temperature dependence of the magnetic moment normalized to the initial value $m_0$ at *T* = 12 K. Since the magnetic moment is recorded at zero field, its changes below $T_c$ can be ascribed to the response of the S layer to the F-layer $m_r$ stray fields. We observe that $m_r$ has the same direction of the applied field before its removal, whereas the F-stray field in the S layer is oriented in the opposite direction; thus, we expect a paramagnetic Meissner effect of the S layer, as sketched in Fig. 2(a). This effect leads to an increase of the measured magnetic moment of a few percent below $T_c$ provided by a sufficient thickness of the F layer, as already observed in S/F heterostructures of different composition [41-43]. In our samples, we observe the following features: (*i*) In the sample with $d_F$ = 3 nm, $m_r$ remains almost unchanged across $T_c$, and only $\frac{1}{m_0}\frac{dm}{dT}$ allows the determination of the $T_c$, suggesting the negligible effect of the F-layer stray field. (*ii*) The samples with $d_F$ = 7 and 10 nm show a clear step in the T dependence of the magnetic moment at ∼ 8.8 K, which is taken as Nb $T_c$ value, in good agreement with the previous reports on Nb/Py multilayers [44,45]. Moreover, in Fig. 2(d), the magnetic moment measured with increasing temperature is lower than the initial one. Indeed, the flux expulsion of the S layers in the superconducting state in response to the stray fields of the F layer may lead to an out-of-plane component in F at the S/F interface and, hence, to a reduction of $m_r$ in the Py layer [44,45].

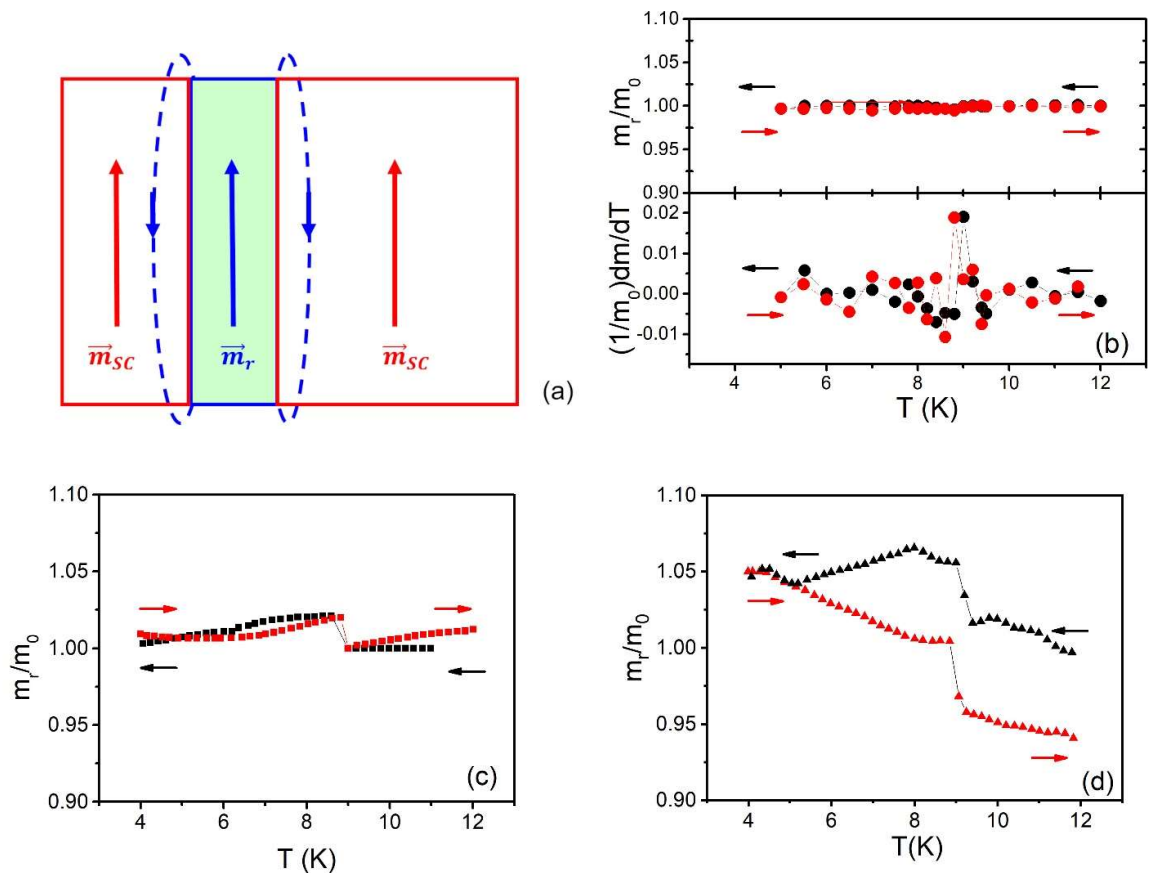

**FIG. 2.** (a) The F-layer stray field lines with respect to both $m_r$ and $m_{SC}$. Panels (b) - (d) report the temperature dependence of the normalized remanent magnetic moment $\frac{m_r}{m_0}$. At $T$ = 12 K, the Nb (30 nm)/Py ($d_F$)/ Nb (400 nm) trilayers were saturated. Then $H$ was removed, and $m_r$ of the trilayers was scanned from 12 to 4 K (down curves: black curves) and then back to 12 K (up curves: red curves): (b) $d_F$ = 3 nm, (c) $d_F$ = 7 nm and (d) $d_F$ = 10 nm.

The *m(T)* dependence of the trilayers has been also investigated in ZFC and FC measurements in the presence of an applied field that leads to the saturation of the F layer, as shown in Fig. 3. The ZFC curves for $d_F$ of 3 and 7 nm, in Figs. 3(a) and 3(b), respectively, are characterized by a decrease of the signal below $T_c$ due to the diamagnetism of the S layers in response to the applied field. For our films, by supposing the 400 nm-thick Nb layer in the Meissner state, the expected $m_{SC}$ would be ∼ -1 μAm², conversely, we observe a diamagnetic signal of ∼ -2 μAm² at $T$ = 4 K. This enhancement of the flux expulsion of a factor of two will be recalled in the following.

By increasing the temperature from 4 K in the ZFC curve, we observe a slight decrease of the diamagnetic contribution at ∼ 6 K, likely due to the transition into the normal state of the 30 nm-thick Nb layer [27]. In Fig. 3(c), the ZFC curve of the trilayer with $d_F$ = 10 nm shows a paramagnetic signal below $T_c$. We can ascribe this behavior to a twofold effect: (i) the thicker F layer contributes to a higher positive offset almost unchanged across

$T_c$ in the ZFC curve and (ii) a paramagnetic Meissner effect contribution due to the F-stray field in the S layers [41].

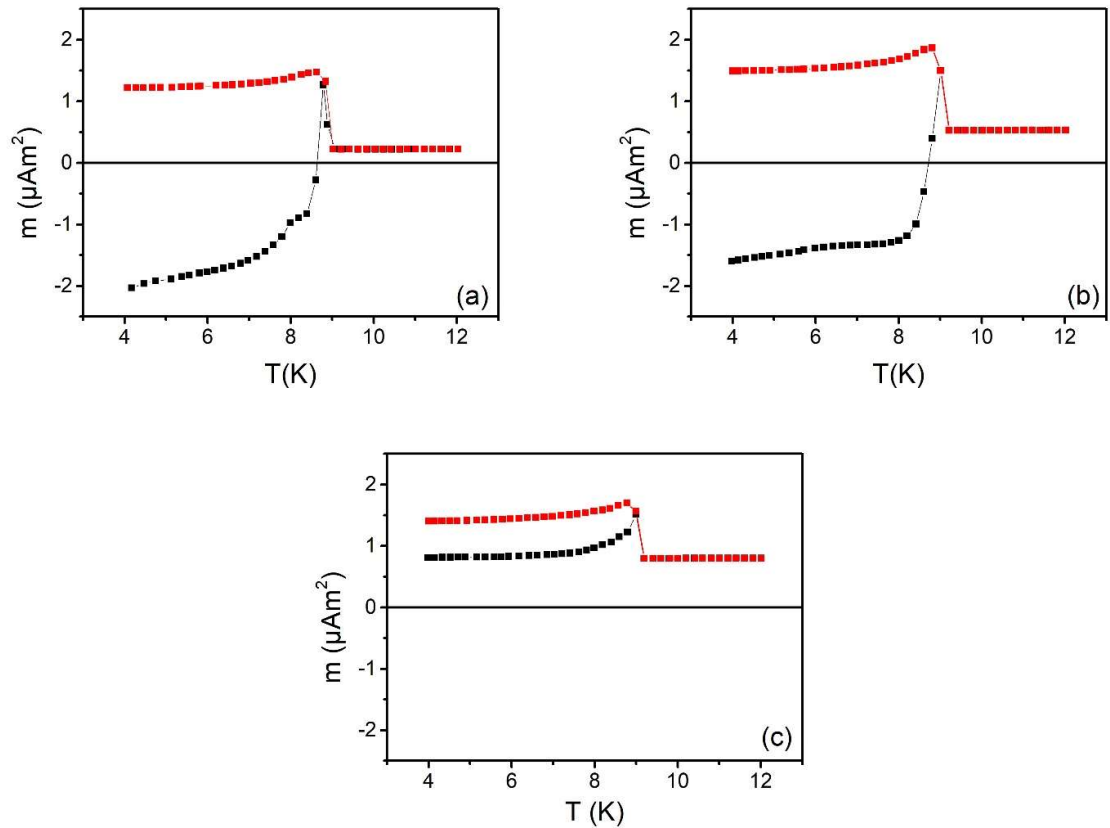

**FIG. 3.** Zero-field cooled (ZFC; black) and field cooled (FC; red) curves of Nb (30nm) /Py ($d_F$) / Nb (400 nm) trilayers by setting an in-plane magnetic field $\mu_0 H_0$ = 30 mT: (a) $d_F$ = 3nm, (b) $d_F$ = 7 nm, and (c) $d_F$ = 10 nm.

At the lower critical temperature $T_{c1}(\mu_0 H_0)$, i.e., at the temperature when the applied field $\mu_0 H_0$ is equal to the lower critical field $\mu_0 H_{c1}(T)$, Abrikosov vortices start penetrating the 400-nm Nb film. In bulk Nb samples, in the ZFC process, the magnitude of the diamagnetism starts to decrease at $T_{c1}(\mu_0 H_0)$ and disappears at the upper critical field $T_{c2}(\mu_0 H_0)$ by increasing the temperature [46,46]. In the FC process, pinning prevents flux lines from being expelled from the sample when the temperature reaches $T_{c1}$ [46,47]. Thus, below such a temperature, the S magnetization saturates to a smaller value than the one in ZFC measurements [48,49]. In our samples, since the paramagnetic signal of the vortex cores adds to the ferromagnetic signal, we clearly observe a paramagnetic peak around $T_c$ in the ZFC measurements and a persistent paramagnetic signal below $T_c$ in the FC measurements of Fig.

3, as already observed in V/Fe bilayers [50]. Following these considerations, from Figs. 3(a) - 3(c), we can state that, for the 400-nm Nb film, $T_{c1}(\mu_0 H_0 = 30$ mT$) \cong 8.0$ K and $T_{c2}(\mu_0 H_0 = 30$ mT$) \cong 8.8$ K.

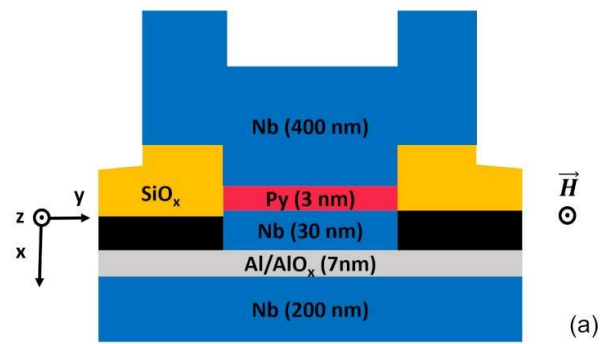

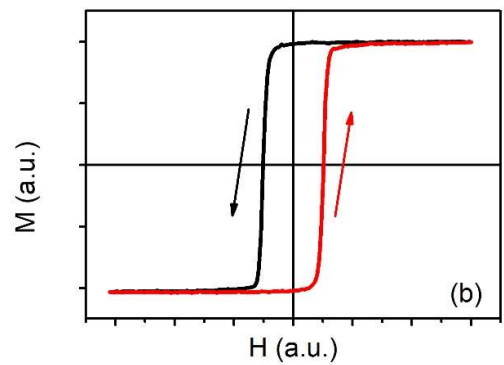

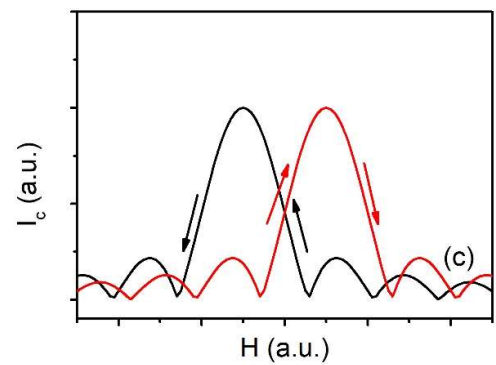

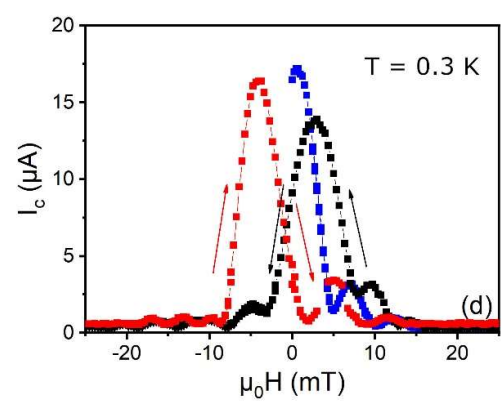

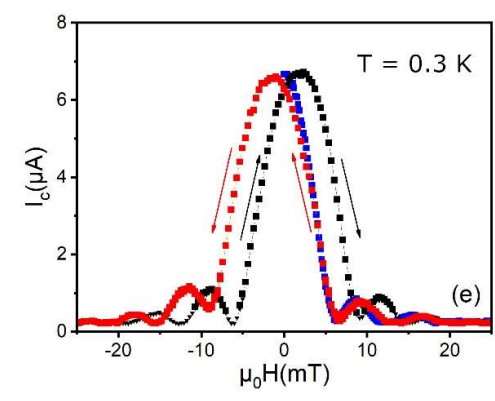

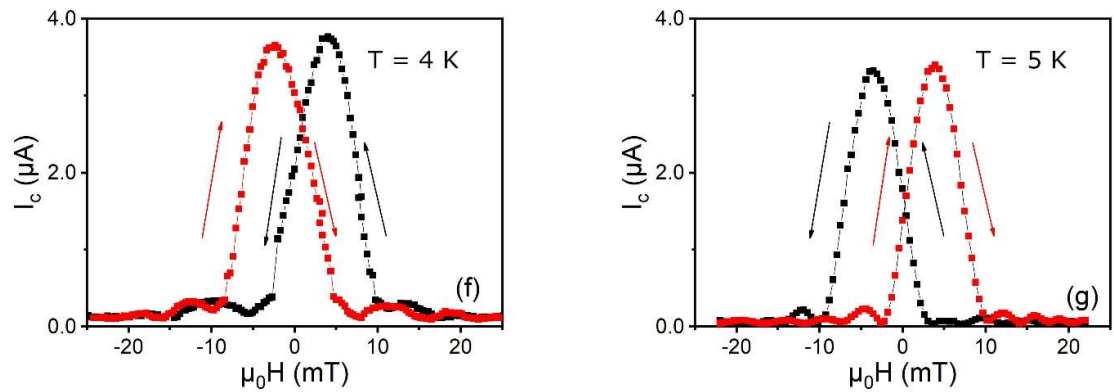

**FIG 4.** (a) Superconductor/insulator/thin superconducting film/ferromagnet/superconductor (SIsFS) Josephson junctions (JJs) sketch; for further details on the fabrication process and the characterization, see Ref. [20]. The magnetic field $H$ is applied along the z direction. (b) For a JJ containing a F-layer characterized by a magnetization curve, (c) the $I_c(H)$ curve is expected to be shifted depending on the sweeping field direction. Experimentally observed $I_c(H)$ dependence (d) for a square 25μm² SIsFS JJ at $T = 4$ K with $d_F = 3$ nm and (e) for a circular 7 μm² JJ with $d_F = 3$ nm at $T = 300$ mK, (f) $T = 4$ K and (g) $T = 5$ K. Arrows indicate the sweeping direction of $H$ for each curve: the blue curves are the curves obtained from the demagnetized state, the black curves are obtained by sweeping $H$ from positive to negative fields, whereas the red curves are obtained from negative to positive fields.

The magnetic characterization of the trilayers is crucial to a comprehensive study of the anomalous inverse magnetic hysteresis of the $I_c$ in the SIsFS JJs observed in Ref. [20] and sketched in Fig. 4(a). It is well known that, in samples containing a F-barrier, to evaluate the total magnetic flux through the junction $\Phi$, the F magnetization flux $\Phi_M$, which is given by $\Phi_M = \mu_0 M_F L d_F$, with $L$ the cross-section width, has to be considered. Hence, the total magnetic flux through the junction is $\Phi = \mu_0 H L d_m + \mu_0 M_F L d_F$, where the thickness of the material penetrated by the applied field is $d_m = 2\lambda_L + d_s + d_F + d_{ox}$, with $d_s$ and $d_{ox}$ the thicknesses of the intermediate Nb, being $< \lambda_L$, and AlO$_x$, respectively [18]. In this case, since the hysteretic dependence of $M_F$ is on $H$, one should observe a hysteresis of the $I_c(H)$ curve depending on the sweeping direction of $H$. The resulting Fraunhofer patterns are shifted in the field to a point where the flux due to the external field cancels out the flux due to the magnetization. Specifically, we expect that, when $H$ is ramped from positive to negative fields [black $I_c(H)$ curves in Fig. 4(b)] the global maximum of the Fraunhofer-like pattern should be shifted toward negative fields because of the positive remanence of the F layer [black $I_c(H)$ curve in Fig. 4(c)], whereas when $H$ is ramped from negative to positive fields [red curves in Fig. 4(b)] it should be shifted toward positive fields [red $I_c(H)$ curve in Fig. 4 (c)]. Thus, in the simple case of a homogeneous F barrier in a single domain state, the Fraunhofer pattern is simply offset by a factor

$$\mu_0 H_{shift} = -\frac{\mu_0 M_S d_F}{(2\lambda_L + d_s + d_F + d_{ox})}, \quad (1)$$

where $M_s$ is the saturation magnetization of the F layer [51-53]. If the F layer does not maintain a uniform magnetization when the applied field is pointing in its opposite direction, the shift field is found to correspond approximately to the coercive field of the F layer [17,18,54-56].

We rather observe and discuss a shift of the $I_c(H)$ curves in the opposite expected direction, as clearly shown for a square 25 μm² JJ in Fig. 4(d) and a circular 7 μm² JJ in Figs. 4(e) and 4(f), with a 3 nm-thick Py interlayer. Evidently, an additional magnetic flux with opposite sign with respect to the one due to $M_F$ must be considered to explain the observed $I_c(H)$ curves. From the results of the magnetic characterization of these trilayers, we can conclude that, for a 3 nm-thick Py layer, the effects of the stray fields are negligible. On the other side, both the inverse and the electromagnetic proximity effects predict an induction field into S, $B_{SC}$, antiparallel to $M_F$. It has been proposed that, neglecting the spin polarization of the S layer at the S/F interface, the electromagnetic proximity effect should induce a shift in the $I_c(H)$ curves for tunnel SIS JJs with one S electrode covered by F [30]. Moreover, the presence of an additional magnetization in S is consistent with the enhancement of the diamagnetic signal observed in the ZFC curves of the Nb (30 nm) /Py (3 nm) / Nb (400 nm) trilayer in Fig. 3(a). Above $T$ = 6 K, the 30 nm-thick Nb layer [27] is in the normal state, and we observe the magnetic response from the bilayer Py (3 nm)/ Nb (400 nm). Hence, we can apply an electromagnetic proximity model proposed for S/F bilayers with no stray fields, $d_S \ggg \lambda_L$, $d_F \ll \lambda_L$ and a uniform in-plane magnetization since, for $\mu_0 H = 30$ mT, the F layer is fully magnetized. The profile of the magnetic induction in S at the S/F interface is

$$B_{SC} = -\mu_0 M_F Q e^{x/\lambda_L}, (2)$$

with

$$Q = \int_0^{d_F} \frac{x'}{\lambda_L^2} dx, (3)$$

where $x$ is the axis orthogonal to the S/F interface where the origin has been set [24]. In our samples, $B_{SC}$ should lead to an enhancement of $m_{SC}$ with respect to the perfect diamagnetic case of ~ -10⁻² μAm², a value that is two orders of magnitude less than to the one experimentally measured (~ -1 μAm²).

We can conclude that the strong enhancement of the diamagnetic signal cannot be justified by the presence of only the electromagnetic proximity effect, but we must also consider the possibility of spin polarization of the S at the interface with the F layer. Recently, it has been theoretically shown that the spin polarization induced in the S itself can lead to a shift of the peak of the $I_c(H)$ curves [29], which is opposite to the displacement by $M_F$, as we have experimentally observed. According to the inverse proximity theory, the magnetization in the S at the interface with F $M_{SC}$ cannot exceed the value of $M_F$ [57]. In our case, $\mu_0 M_F$ is ~ 1T; hence, the magnetic moment detected is consistent if we at least assume a value of $\xi_s$ of the order of 10 nm, as reported in the literature for a Nb polycrystalline thin film [57] and for Nb/Py systems [27].

Therefore, if we suppose that, in the SIsFS JJs, we are in the presence of a magnetization $M_{sc}$ induced by inverse proximity effect in two S layers adjacent to the F, and we assume that the two S/F interfaces are identical, neglecting the electromagnetic proximity effect and the stray field, the total magnetic flux through the junction is [29]

$$\Phi = \mu_0 H\, d_m L + \mu_0 M_F(H) d_F L\, [1 - \gamma], \quad (4)$$

where $\gamma$ is a parameter that considers the contribution to the total flux through the junction caused by the spin polarization in the S:

$$\gamma = \left| \frac{M_{sc}}{M_F} \frac{2\xi_s}{d_F} \right|. \quad (5)$$

To observe the inverse magnetic hysteresis as reported in Figs. 4(d) - 4(f), $\gamma$ must be larger than unity [29]. In our samples, $\gamma$ is ~ 1.11. Hence, assuming a value of 10 nm for $\xi_s$ [58], the observed shift of the global maximum $\mu_0 H_{shift} \approx 3$ mT ($\mu_0 M_F \sim 1$ T) can be justified by a value of $|\mu_0 M_{sc}|$ at most equal to 0.16 T, which is consistent with the inverse proximity theory being $|\mu_0 M_{sc}| < \mu_0 M_F$ [57].

The formula of the flux has been expressed using the parameter $\gamma$ to point out the crucial role of the ratio $\frac{2\xi_s}{d_F}$ to observe this phenomenon; the F thickness must be sufficiently small compared with $2\xi_s$. The use of dilute ferromagnetic alloys, e.g., PdFe, PdNi, and CuNi, characterized by low value of the exchange energy, on one hand, has allowed to prepare thin film SFS sandwiches with homogeneous ferromagnetic interlayers with thickness of tens of nanometers, i.e., of the order of $\xi_f$, but on the other hand, may prevent the observation of this unconventional effect [17,18,51,59]. Moreover, the domain structure of these ferromagnetic interlayers for large-area junctions is quite small scale; hence, the magnetization can be completely averaged at the scale of the F layer, leading to nonhysteretic $I_c(H)$ curves [11,60-62]. Since the use of a strong ferromagnet can suppress the value of the critical current of the JJs, commonly, thickness of the F layer (order of nanometers) or pseudo-spin valve structure [63-65] has been employed smaller. In the latter structures, the use of a metallic buffer layer can compromise the possibility to induce the spin polarization in the S layers. Furthermore, buffer layers have been employed also for JJs containing a single F layer to avoid the formation of magnetic dead layer [66].

Therefore, the choice of the materials and the geometry of the device, along with the characteristic length scales of the S and F layers, and particularly the ratio $\frac{\xi_s}{d_F}$, play a crucial role in the observation of the inverse magnetic hysteresis.

Finally, to gain insights on the effect of the spin polarization on the Josephson magnetometry, we decided to perform the $I_c(H)$ measurements for a circular 7 μm$^2$ JJ with $d_F$ = 3 nm at different temperatures, see Figs. 4(e)-4(g). As shown in Fig. 4(g), at $T$ = 5 K, the ordinary behavior is restored, i.e., we observe the maximum of the down curve at a negative field and the maximum of the up curve at a positive field. The measurement at $T$ = 5 K allow us to rule out the possibility that this uncommon behavior can be due to the F-stray fields. Indeed, in this narrow temperature range, it is very unlikely that the change of the behavior from an inverse to an ordinary hysteresis can be ascribed to changes in the domain structure of the ferromagnet. Conversely, we expect a strong dependence of the magnetization induced in the superconductor $M_{SC}$ on the temperature, as predicted in Ref. [12] and experimentally observed [67]. It means that, at $T$ = 5 K, the parameter $\gamma$ becomes less than unity because of the decreasing of the induced magnetization.

## IV. CONCLUSIONS

We have investigated the magnetic properties of S/F/S heterostructures based on thin films of Nb and Py. By monitoring $m(T)$ above and below $T_c$ of the heterostructure, we have observed that the magnetic properties of the structures depend on the thickness of the F interlayer. Our measurements clearly point out that in trilayers with F-layer thickness > 7 nm, the effects of the F-stray fields become relevant: the sample with a 10-nm Py layer shows a decrease of a factor of six of the diamagnetic signal in the hysteresis loop at $T$ = 4 K with respect to the two thinner ones and the presence of the paramagnetic Meissner effect [41] in the thermoremanence measurements below $T_c$. Moreover, for all the samples, we clearly observed a paramagnetic peak around $T_c$ in the ZFC curves that has been related to vortices in the mixed state and a paramagnetic signal in the FC curves below $T_{c1}$ due to pinning-induced flux trapping. These outcomes give a further insight on the magnetic properties of the SFS heterostructures, consistent with the anomalous inverse magnetic hysteresis of the critical current $I_c$ in the SIsFS JJs observed in Ref. [20]. The shift of the $I_c(H)$ curves in the opposite direction of the F magnetization is related to spin-polarization phenomena at the S/F interface and to the inverse proximity effect [29], which is typically investigated by a highly sensitive local magnetic probe, e. g., polarized neutron reflectometry [68]. Thus, we show that the magnetic characterization carried out in this paper is crucial to properly design S/F heterostructures as memory cells or for spintronic devices.

## ACKNOWLEDGMENTS

The authors are grateful to E. Di Gennaro for his assistance in x-ray diffraction analysis.

## APPENDIX

A structural and morphological analysis of our thin film surfaces has been performed by atomic force microscopy (AFM). In Fig. 5, we show some AFM images (2 x 2 μm$^2$) obtained for a 40-nm thick Nb layer (a) before the etching cleaning, (b) after, and (c) for a Nb/Py($d_F$=3nm) bilayer. The mean-square-root roughness is $R_q \approx 0.4$ nm for the Nb film before the etching, whereas $R_q \approx 0.5$ nm for the Nb film after the cleaning process and for the 3 nm-thick Py grown on the etched Nb. This result suggests that the etching process does not significantly affect the S/F interface and that the overall roughness of the S/F junction interface is determined by the roughness of the Nb interlayer in the junction. The grain size is so fine that it cannot be appreciated with our image software.

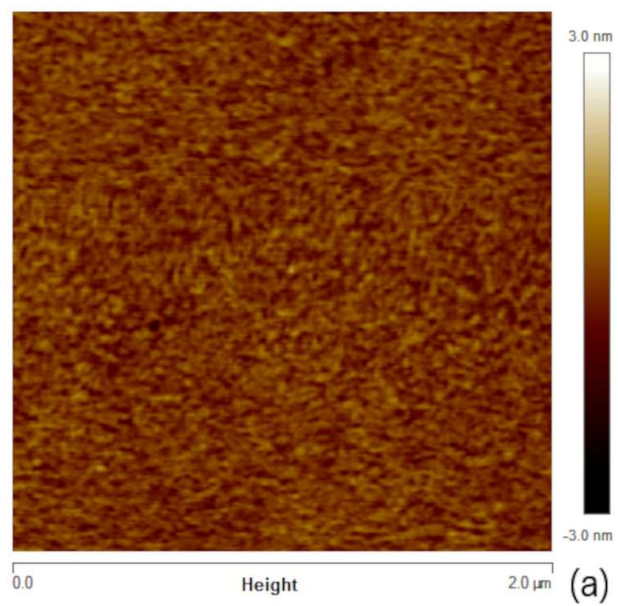

(a)

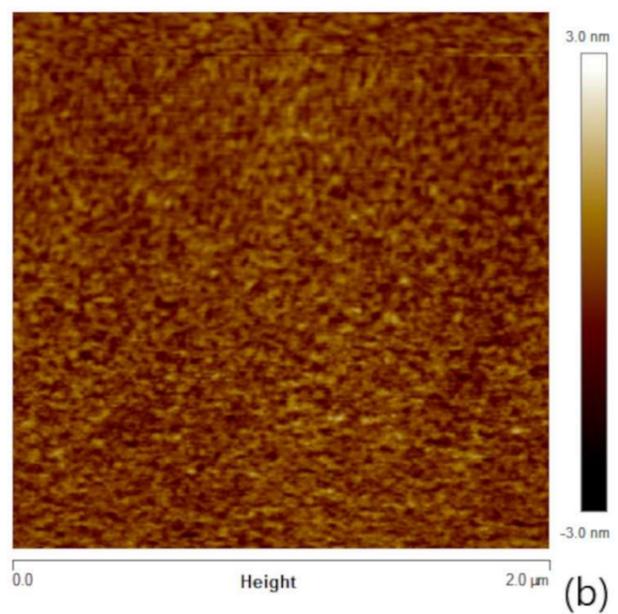

(b)

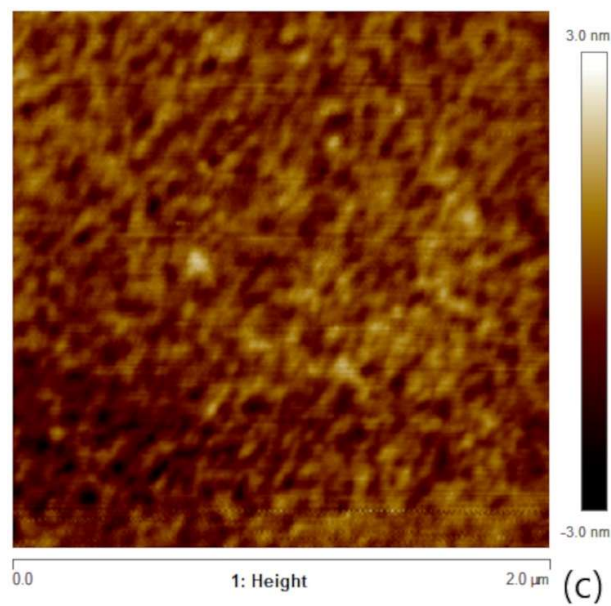

**FIG. 5**: Topography maps (2 x 2 µm$^2$) obtained by atomic force microscopy for a 40 nm-thick Nb layer (a) before the etching cleaning, (b) after, and (c) for a Nb/Py($d_F$=3nm) bilayer.